\documentclass[hidelinks,a4,12pt]{article}
\usepackage{amsmath}
\usepackage{amsthm}
\usepackage{amssymb}
\usepackage{amscd}
\usepackage{mathrsfs}
\usepackage{feynmf}
\usepackage[pdftex]{graphicx}
\usepackage[pdftex,bookmarks=true,pdfauthor={},pdftitle={},pagebackref=false]{hyperref}
\usepackage{youngtab}
\usepackage{cite}
\usepackage{color}
\usepackage{mdframed}
\usepackage{breakurl}
\usepackage{slashed,bbold}
\usepackage{hyperref} 
\usepackage[hang,flushmargin]{footmisc}
\usepackage{setspace}
\usepackage{accents}
\hypersetup{colorlinks,urlcolor=blue,citecolor=blue,linkcolor=blue}

\hypersetup{colorlinks,urlcolor=blue,citecolor=blue,linkcolor=blue}

\numberwithin{equation}{section}

\def\oomega{\accentset{\circ}{\omega}}

\def\onabla{\accentset{\circ}{\nabla}}
\def\notrQ{\mathcal{Q}\hspace*{-0.5cm}\rotatebox{45}{\raisebox{-0.25cm}{$\boldsymbol{\longrightarrow}$}}\!}
\def\notrR{\mathcal{R}\hspace*{-0.53cm}\rotatebox{45}{\raisebox{-0.25cm}{$\longrightarrow$}}\!}
\def\notrT{T\hspace*{-0.45cm}\rotatebox{45}{\raisebox{-0.20cm}{$\longrightarrow$}}\!\!}

\def\notrQbf{\,\boldsymbol{Q}\hspace*{-0.59cm}\rotatebox{45}{\raisebox{-0.30cm}{$\boldsymbol{\longrightarrow}$}}\!}
\def\notrmu{\,\mu\hspace*{-0.51cm}\rotatebox{45}{\raisebox{-0.32cm}{$\longrightarrow$}}\!}
\def\notrB{\boldsymbol{B}\hspace*{-0.59cm}\rotatebox{45}{\raisebox{-0.30cm}{$\boldsymbol{\longrightarrow}$}}\!}
\def\notrE{E\hspace*{-0.5cm}\rotatebox{45}{\raisebox{-0.25cm}{$\boldsymbol{\longrightarrow}$}}\!}

\def\notrT{\mathscr{T}\hspace*{-0.62cm}\rotatebox{45}{\raisebox{-0.28cm}{$\boldsymbol{\longrightarrow}$}}\!}
\def\notrDelta{\Delta\hspace*{-0.56cm}\rotatebox{45}{\raisebox{-0.28cm}{$\boldsymbol{\longrightarrow}$}}\!}

\allowdisplaybreaks

\begin{document}

\setlength{\oddsidemargin}{0cm}
\setlength{\baselineskip}{7mm}

\thispagestyle{empty}
\setcounter{page}{0}
\onehalfspacing

\begin{flushright}

\end{flushright}

\vspace*{-1cm}

\begin{center}
{\bf \Large }
\end{center}

\vspace*{2cm}

\begin{center}
{\bf \Large Chiral Transport in Metric-Affine Geometries}


\vspace*{1cm}

Miguel \'A. V\'azquez-Mozo\footnote{\tt 
vazquez@usal.es}

\end{center}

\begin{center}
{\sl Departamento de F\'{\i}sica Fundamental, Universidad de Salamanca \\ 
 Plaza de la Merced s/n,
 37008 Salamanca, Spain
  }
\end{center}

\vspace*{1cm}

\centerline{\bf \large Abstract}

\vspace*{0.25cm}

\noindent
Anomalous 
transport in equilibrium fermionic fluids chirally coupled to background Weyl-type nonmetricity 
is studied. A formal descent analysis is 
carried out in which the dependence of the 
anomaly polynomial on the nonmetricity tensor is encoded in a Weyl
invariant four-form. The constitutive relation of the axial-vector current
is evaluated from the equilibrium partition function obtained using transgression techniques, 
showing
the existence of nonmetricity-mediated chiral separation effects driven
by the fluid's vorticity and the Weyl magnetic field. A second nonminimal coupling 
of fermionic matter to metric-affine geometries proposed in the literature is also discussed.

\newpage

\tableofcontents

\setcounter{footnote}{0}

\section{Introduction}

The formal study of fluids in nontrivial 
background geometries~\cite{Rezzolla:2013,Andersson:2020phh,Disconzi:2023rtt} 
has found interesting applications to
condensed matter systems (see, e.g.,~\cite{Geracie:2015xfa,Lucas:2017idv,Chernodub:2021nff}). 
At or near equilibrium, the metric components act as sources for
the fluid's velocity, vorticity, acceleration, and local temperature~\cite{Banerjee:2012iz}.
Torsion and curvature, on the other hand, emulate the long-wavelength
effects of lattice dislocations and 
disclinations, respectively~\cite{Kondo1952,Bilby:1955,Katanaev:1992kh,Ruggiero:2003tw,Hehl:2007bn}. 
Nonmetricity~\cite{Schouten1924} 
(see, e.g.,~\cite{Hehl:1994ue,Ponomarev2017,JimenezCano:2021rlu} for reviews) has also been added
to this picture. It was pointed out in ref.~\cite{Palumbo:2024vqv} that  
certain type of background nonmetricity  
could describe the long-distance effect of lattice point defects in Weyl semimetals, 
providing an interpretation for the phenomena described in~\cite{Pires:2022fdf} and affecting chiral transport 
in these systems. More recently, momentum space nonmetricity has 
been associated to nonlinear odd acoustoelastic effects~\cite{Jain:2026fbu}.

The aim of this paper is to provide a detailed analysis of
chiral transport driven by background spacetime
nonmetricity in relativistic fermionic fluids at equilibrium. We will consider
in the first place the chiral coupling
proposed in~\cite{Palumbo:2024vqv} and apply
transgression methods, which have been successful in various physical 
situations (see, e.g.,~\cite{Jensen:2012kj,Jensen:2013kka,Jensen:2013rga,Manes:2018llx,Valle:2021nfv}),
to compute the equilibrium partition function.
To implement the descent analysis,
we construct an invariant that captures the nonmetricity dependence
in the anomaly polynomial, reproducing the axial 
anomaly for ``pure Weyl'' nonmetricity. The constitutive relation for the chiral current  
derived from
the equilibrium partition function shows 
the existence of nonmetricity-induced chiral separations effects
mediated by the fluid's vorticity and the magnetic Weyl field. 
These results 
are also applied to 
the nonminimal coupling built in~\cite{Rigouzzo:2023sbb}, where both chiral separation effects
are driven by additional nonmetricity components besides the Weyl gauge field, as well as torsion.

We open our discussion in section~\ref{sec:setup} with a review 
of the issue of the coupling of fermionic matter
to nonmetricity. In section~\ref{sec:NYlike_inv} a nonmetricity invariant
is constructed, analogous to the torsional Nieh-Yan invariant, 
in terms of which the anomaly 
polynomial is written in section~\ref{sec:anomaly_pol}. 
Section~\ref{sec:constitutive_relations} contains the analysis of the
nonmetricity-induced terms in the constitutive relation
for the axial-vector current.
A brief discussion of the results is offered in section~\ref{sec:closing}. To make the exposition
self-contained, basic aspects of nonmetricity are presented in
Appendix~\ref{sec:nonmetricity_generalities}, while
generalized metric rescalings are discussed in Appendix~\ref{sec:appendix}.

\section{Coupling fermionic matter to nonmetricity}
\label{sec:setup}

Nonmetricity measures the incompatibility of a connection with the 
metric,~$\nabla_{\mu}g_{\alpha\beta}\neq 0$
(see, e.g,~\cite{Schouten1924,Hehl:1994ue,Ponomarev2017,JimenezCano:2021rlu}, as
well as Appendix~\ref{sec:nonmetricity_generalities}). 
In the Cartan formalism, this is encoded in the one-form nonmetricity  
tensor\footnote{Throughout the paper, frame indices are denoted by Latin fonts. 
Following a standard notation in the literature (see, e.g.,~\cite{Hehl:1994ue}), 
the differential forms interior product is 
indicated by~$V\rfloor \zeta\equiv \imath_{V}\zeta$~\cite{Nakahara:2003nw}, 
with~$V$ a tangent space vector
and~$\zeta$ a $p$-form. 
Tangent and cotangent space bases are respectively represented by~$\{e_{a}\}$ and~$\{\theta^{a}\}$ 
(with~$a=0,\ldots,3$) and they satisfy~$e_{b}\rfloor\theta^{a} =\delta^{a}_{b}$.
Spacetime indices are indicated by Greek fonts. 
} 
\begin{align}
\mathcal{Q}_{ab}\equiv -Dg_{ab},
\end{align}
where~$g_{ab}$ is the frame metric,~$ds^{2}=g_{ab}\theta^{a}\otimes\theta^{b}$,
and~$D$ denotes the covariant exterior differential
\begin{align}
D\mathfrak{T}^{a_{1}\ldots a_{p}}_{\hspace*{0.8cm}b_{1}\ldots b_{q}}&\equiv
d\mathfrak{T}^{a_{1}\ldots a_{p}}_{\hspace*{0.8cm}b_{1}\ldots b_{q}}
+\omega^{a_{1}}_{\,\,\,\,c}\wedge\mathfrak{T}^{c\ldots a_{p}}_{\hspace*{0.7cm}b_{1}\ldots b_{q}}
+\ldots+\omega^{a_{p}}_{\,\,\,\,c}\wedge\mathfrak{T}^{a_{1}\ldots c}_{\hspace*{0.7cm}b_{1}\ldots b_{q}}
 \nonumber \\[0.2cm]
&+(-1)^{r+1}\mathfrak{T}^{a_{1}\ldots a_{p}}_{\hspace*{0.8cm}c\ldots b_{q}}\wedge 
\omega^{c}_{\,\,\,b_{1}}+\ldots
+(-1)^{r+1}\mathfrak{T}^{a_{1}\ldots a_{p}}_{\hspace*{0.8cm}b_{1}\ldots c}\wedge \omega^{c}_{\,\,\,b_{q}},
\label{eq:covariant_dif1}
\end{align}
with~$\mathfrak{T}^{a_{1}\ldots a_{p}}_{\hspace*{0.8cm}b_{1}\ldots b_{q}}$ 
a generic  $r$-form tensor of type~$(p,q)$ and~$\omega^{a}_{\,\,\,b}$ the
spin connection. 
In what follows, we will use the orthonormal gauge~\cite{Hehl:1994ue}
and choose the cotangent basis such 
that~$g_{ab}=\eta_{ab}\equiv\mbox{diag}(-1,1,1,1)$.
Since~$d\eta_{ab}=0$,
the nonmetricity one-form is then fully determined by the symmetric part of the spin 
connection
\begin{align}
\mathcal{Q}_{ab}\equiv -D\eta_{ab}=2\omega_{(ab)}.
\label{eq:nonmetricity_def}
\end{align}
A further advantage of 
the orthonormal gauge is that the Dirac matrices~$\gamma^{a}$ satisfy the standard Lorentzian
Clifford algebra~$\gamma^{(a}\gamma^{b)}=-\eta^{ab}\mathbb{1}$, with~$d\gamma^{a}=0$. 

The nonmetricity tensor defined in~\eqref{eq:nonmetricity_def}
can be decomposed into its trace and shear (traceless) pieces
\begin{align}
\mathcal{Q}_{ab}&\equiv\notrQ_{ab}+\mathcal{Q}\eta_{ab},
\label{eq:Q_Qbar_def}
\end{align}
both transforming irreducibly under~$\mbox{SO(1,3)}$. Here and in
the following, we use the notation of ref.~\cite{Hehl:1994ue} to denote the traceless
part of an arbitrary rank-two tensor~$\mathscr{T}_{ab}$
\begin{align}
\notrT_{\!ab}\equiv \mathscr{T}_{ab}-{1\over 4}\eta_{ab}\mathscr{T}^{c}_{\,\,\,\,\,c}.
\label{eq:Rbar_def}
\end{align}
In addition to the Weyl one-form~$\mathcal{Q}$, 
a second vector field can be defined from the components of the nonmetricity tensor by
\begin{align}
\widehat{\mathcal{Q}}_{a}&\equiv e_{b}\rfloor\mathcal{Q}^{b}_{\,\,\,a} \nonumber \\[0.2cm]
&=e_{b}\rfloor\notrQ^{b}_{\,\,\,a}+e_{a}\rfloor\mathcal{Q}.
\label{eq:second_nm_gauge_field}
\end{align}
The associated one-form is given by~$\widehat{\mathcal{Q}}\equiv \widehat{\mathcal{Q}}_{a}\theta^{a}$.

Under Weyl transformations\footnote{See Appendix~\ref{sec:appendix} for the relevant 
identities
and terminology.}, the 
components of the nonmetricity tensor change as (see, e.g., \cite{Hehl:1994ue,Iosifidis:2018zwo})
\begin{align}
\mathcal{Q}&\longrightarrow \mathcal{Q}-2d\sigma, \nonumber \\[0.2cm]
\notrQ_{ab}&\longrightarrow \notrQ_{ab}.
\label{eq:Weyl_transf_def}
\end{align}
The transformation of the trace one-form~$\mathcal{Q}$ explains 
it being dubbed the Weyl gauge field. 
Given the Weyl 
invariance of~$\notrQ^{a}_{\,\,\,b}$ and eq.~\eqref{eq:transform_frame_e},
we find for the second nonmetricity vector field 
\begin{align}
\widehat{\mathcal{Q}}_{a}\longrightarrow e^{-\Lambda\sigma}\big(\widehat{\mathcal{Q}}_{a}
-2 e_{a}\rfloor d\sigma\big),
\end{align}
so~$\widehat{\mathcal{Q}}$ 
and~$\mathcal{Q}$ undergo the same gauge transformation [see the first 
identity in~\eqref{eq:general_transform_frame}]. 

After these basic preliminaries, we proceed to review the coupling of
fermionic matter to nonmetricity. Lorentz spinors couple minimally to background 
geometry through the 
action (see, e.g.,~\cite{Freedman:2012zz})
\begin{align}
S=-{1\over 2}\int d^{4}x\,e\left[i\overline{\psi}\gamma^{a}\nabla_{a}\psi
+\Big(i\overline{\psi}\gamma^{a}\nabla_{a}\psi\Big)^{\dagger}\right],
\label{eqw:action_mincoupgrav}
\end{align}
where~$\gamma^{a}$ are the frame Dirac matrices satisfying~$\gamma^{(a}\gamma^{b)}=-\eta^{ab}\mathbb{1}$,
and~$e$ is the determinant of the vierbein, whose components~$e_{\,\,\,a}^{\mu}$
are defined by~$e_{a}\equiv e^{\mu}_{\,\,\,a}\partial_{\mu}$.
The fermion covariant derivative is given by
\begin{align}
\nabla_{a}\psi=e_{\,\,\,a}^{\mu}\big(\partial_{\mu}+\Gamma_{\mu}\big)\psi,
\end{align}
with~$\Gamma_{\mu}$ the Fock-Ivanenko 
coefficients~\cite{Fock1929_1,Fock1929_2}
\begin{align}
\Gamma_{\mu}&=-{1\over 4}\omega_{\mu ab}\gamma^{[a}\gamma^{b]},
\label{eq:FI_minimal_coefficients}
\end{align}
expressed in terms of the spin 
connection components,~$\omega^{a}_{\,\,\,b}\equiv\omega_{\mu\,\,\,b}^{\,\,\,a}dx^{\mu}$. 

However, Lorentz spinors do not minimally couple to 
nonmetricity~\cite{BeltranJimenez:2020sih}. A simple way to see this is by writing the full
spin connection as
\begin{align}
\omega^{a}_{\,\,\,b}&=\widetilde{\omega}^{a}_{\,\,\,b}+{1\over 2}\mathcal{Q}^{a}_{\,\,\,b},
\label{eq:omega=omegatilde+1/2Q}
\end{align}
with~$\widetilde{\omega}_{ab}=\omega_{[ab]}$ the antisymmetric part of the
spin connection [see eq.~\eqref{eq:omega_tilde_def} and discussion therein].
Due to antisymmetrization, the second term on the right-hand side of the previous equation 
does not contribute to the 
Fock-Ivanenko coefficients~\eqref{eq:FI_minimal_coefficients}, 
so we have~$\nabla_{a}=\widetilde{\nabla}_{a}$ and the fermion action~\eqref{eqw:action_mincoupgrav} 
is then recast as
\begin{align}
S&=-{1\over 2}\int d^{4}x\,e\left[i\overline{\psi}\gamma^{a}\widetilde{\nabla}_{a}\psi
+\Big(i\overline{\psi}\gamma^{a}\widetilde{\nabla}_{a}\psi\Big)^{\dagger}\right] \nonumber \\[0.2cm]
&=-{1\over 2}\int d^{4}x\,e\,
\Big[i\overline{\psi}\gamma^{a}\onabla_{a}\psi
+\Big(i\overline{\psi}\gamma^{a}\onabla_{a}\psi\Big)^{\dagger}
+2\overline{\psi}\gamma_{5}\gamma^{a}\psi \widetilde{\mathcal{S}}_{a}\Big],
\label{eq:fermion_action_minimal2}
\end{align}
with~$\widetilde{\nabla}_{a}$ and~$\onabla_{a}$ the spinor covariant 
derivatives respectively associated
to~$\widetilde{\omega}^{a}_{\,\,\,b}$ and the Levi-Civita spin connection~$\oomega^{a}_{\,\,\,b}$.
To write the expression in the second line we used the result that
torsion couples minimally to fermions through the effective axial-vector 
field~\cite{Datta:1971ic,Datta:1971id,Hehl:1971qi,Hehl:1976kj}
\begin{align}
\widetilde{\mathcal{S}}_{a}=-{1\over 8}\epsilon_{abcd}\widetilde{T}^{bcd},
\end{align}
where~$\widetilde{T}^{a}=d\theta^{a}+\widetilde{\omega}^{a}_{\,\,\,b}\wedge\theta^{b}
\equiv{1\over 2}\widetilde{T}_{bc}^{\,\,\,\,a}\theta^{b}\wedge\theta^{c}$.
Substituting the explicit expression of the components~$\widetilde{T}_{ab}^{\,\,\,\,c}$
given in~\eqref{eq:Ttilde_2form} and using~$\mathcal{Q}_{a[bc]}=0$, we see
that~$\widetilde{\mathcal{S}}_{a}$ is independent of the nonmetricity tensor. 
As a consequence,~$\mathcal{Q}_{abc}$ drops out of the fermion 
action~\eqref{eqw:action_mincoupgrav}, thus proving the absence of minimal coupling 
to background nonmetricity.

Various proposals beyond the minimal prescription~\eqref{eq:FI_minimal_coefficients}
to couple fermions to nonmetricity
have been put forward in the literature. 
One of them postulates a Kosmann coupling~\cite{Kosmann:1971ugf,Janssen:2018exh}
\begin{align}
\Gamma_{\mu}&=-{1\over 4}\omega_{\mu ab}\gamma^{a}\gamma^{b} \nonumber \\[0.2cm]
&=-{1\over 4}\widetilde{\omega}_{\mu ab}\gamma^{[a}\gamma^{b]}
+{1\over 2}\mathcal{Q}_{\mu},
\label{eq:Kosmann_coupling}
\end{align}
resulting in the Weyl gauge field coupling to the fermion vector current. Alternatively, 
axial-vector nonminimal couplings were 
considered in~\cite{Rigouzzo:2023sbb}, where the following action was proposed
\begin{align}
S&=-{1\over 2}\int d^{4}x\,e\left\{i\overline{\psi}\big(1-i\alpha-i\beta\gamma_{5}\big)
\gamma^{a}\nabla_{a}\psi
+\Big[i\overline{\psi}\big(1-i\alpha-i\beta\gamma_{5}\big)
\gamma^{a}\nabla_{a}\psi\Big]^{\dagger}\right\} \nonumber \\[0.2cm]
&=-{1\over 2}\int d^{4}x\,e\left\{i\left[\overline{\psi}\gamma^{a}\onabla_{a}\psi
-\Big(\overline{\psi}\gamma^{a}\onabla_{a}\psi\Big)^{\dagger}\right]
+2\overline{\psi}\gamma_{5}\gamma^{a}\psi\mathcal{S}_{a}\right.
\label{eq:action_nonminimal_rigouzzo_zell} \\[0.2cm]
&\left.+{\alpha\over 2}\big(2\mathcal{T}_{a}
+\widehat{\mathcal{Q}}_{a}-4\mathcal{Q}_{a}\big)\overline{\psi}\gamma^{a}\psi
+{\beta\over 2}\big(2\mathcal{T}_{a}
+\widehat{\mathcal{Q}}_{a}-4\mathcal{Q}_{a}\big)\overline{\psi}\gamma_{5}\gamma^{a}\psi
\right\}.
\nonumber
\end{align}
Here,~$\alpha,\beta$ are real constants and~$\nabla_{a}$ is the fermion covariant derivative
built from the standard
Fock-Ivanenko coefficients~\eqref{eq:FI_minimal_coefficients}.
Unlike in the Kosmann case, here fermions do couple chirally to both nonmetricity gauge 
fields,~$\mathcal{Q}_{a}$ and~$\widehat{\mathcal{Q}}_{a}$, as well
as to the edge torsion~$\mathcal{T}_{a}\equiv T_{ab}^{\,\,\,\,\,\,b}$, in addition 
to the minimal coupling to the screw torsion gauge 
field~$\mathcal{S}_{a}=-{1\over 8}\epsilon_{abcd}T^{bcd}$ 
(cf.~\cite{Imaki:2020csc}). The effective gauge field coupling to the 
axial-vector fermion current reads
\begin{align}
\mathcal{B}_{a}&\equiv {\beta\over 4}(2\mathcal{T}_{a}+\widehat{\mathcal{Q}}_{a}-4\mathcal{Q}_{a}) \nonumber \\[0.2cm]
&={\beta\over 2}\widetilde{\mathcal{T}}_{a},
\label{eq:Bagaugefield}
\end{align}
with~$\widetilde{\mathcal{T}}_{a}\equiv\widetilde{T}_{ab}^{\,\,\,\,\,\,b}$ the  
edge torsion associated 
with the antisymmetric part of the connection,~$\widetilde{\omega}^{a}_{\,\,\,a}$ 
[see eq.~\eqref{eq:Ttilde_2form}].
Switching to spacetime indices, Weyl transformations act by
\begin{align}
\mathcal{B}_{\mu}&\longrightarrow \mathcal{B}_{\mu}+{3\beta\over 2}\partial_{\mu}\sigma.
\end{align}
Analogous expressions can be written for the effective
gauge field coupling to the vector current, with~$\beta$
replaced by~$\alpha$.
The full action remains invariant if fermions transform 
as~$\psi\rightarrow \exp[-{3i\over 2}(\alpha\mathbb{1}+\beta\gamma_{5})\sigma]\psi$.
In fact, expressing the action~\eqref{eq:action_nonminimal_rigouzzo_zell} in the form
\begin{align}
S&=-{1\over 2}\int d^{4}x\,e\left\{i\left[\overline{\psi}\gamma^{a}\widetilde{\nabla}_{a}\psi
-\Big(\overline{\psi}\gamma^{a}\widetilde{\nabla}_{a}\psi\Big)^{\dagger}\right]
+\alpha\widetilde{\mathcal{T}}_{a}\overline{\psi}\gamma^{a}\psi
+\beta\widetilde{\mathcal{T}}_{a}\overline{\psi}\gamma_{5}\gamma^{a}\psi
\right\},
\end{align}
the theory is mapped into one of fermions on a Riemann-Cartan background with 
the vector and axial-vector fermion currents coupled to the edge torsion. The dependence
on the original nonmetricity tensor components is encoded in the effective background 
torsion~$\widetilde{T}^{a}$. 

An alternative chiral coupling of fermions to background nonmetricity 
was built in~\cite{Palumbo:2024vqv}, based on matter fermions transforming 
in the finite-dimensional spinor
representation of~$\mbox{SU(2,2)}$, the double covering of the four-dimensional
conformal group~$\mbox{SO(2,4)}$. 
Fermions then couple to the background through the term~$\overline{\psi}\gamma^{a}\mathcal{V}_{a}\psi$,
with
\begin{align}
\mathcal{V}&\equiv \mathcal{V}_{a}\theta^{a}=\theta^{a}P_{a}-{1\over 2}\omega^{ab}J_{ab}
+\lambda^{a}K_{a}+\mathcal{Q}D,
\end{align}
where~$\lambda^{a}$ is the special 
conformal gauge one-form~\cite{Kaku:1977pa,Kaku:1978nz,Gegenberg:2015gma} and~$P_{a}$,~$J_{ab}$,~$K_{a}$, 
and~$D$ are, respectively, the generators of
spacetime translations, Lorentz transformations, special conformal transformation,
and dilations. Since~$D=-{1\over 2}\gamma_{5}$, this leads to 
the following chiral coupling of fermions to the Weyl gauge field 
\begin{align}
S&=-{1\over 2}\int d^{4}x\,e\,
\Big[i\overline{\psi}\gamma^{a}\onabla_{a}\psi
+\Big(i\overline{\psi}\gamma^{a}\onabla_{a}\psi\Big)^{\dagger}
+2\overline{\psi}\gamma_{5}\gamma^{a}\psi \mathcal{Q}_{a}\Big].
\label{eq:action_palumbo}
\end{align}
Once again,~$\onabla_{a}$ represents the Levi-Civita fermion covariant derivative.
This action remains invariant under 
Weyl rescalings~\eqref{eq:Weyl_transf_def} provided
fermions transform according to~$\psi\rightarrow \exp(2i\gamma_{5}\sigma)\psi$.

\section{A nonmetricity invariant}
\label{sec:NYlike_inv}

We consider the~$\mbox{SO(1,3)}$-invariant three-form
\begin{align}
\mathcal{W}_{3}&\equiv {1\over 2}\mathcal{Q}^{a}_{\,\,\,b}\wedge \mathcal{R}^{b}_{\,\,\,a},
\label{eq:W3d}
\end{align}
where~$\mathcal{R}^{a}_{\,\,\,b}= d\omega^{a}_{\,\,\,b}
+\omega^{a}_{\,\,\,c}\wedge\omega^{c}_{\,\,\,b}$ is the curvature two-form.
This defines a bona fide nonmetricity Chern-Simons form (cf.~\cite{Hehl:1990ir}), shifting 
under Weyl rescalings by a total derivative
\begin{align}
\mathcal{W}_{3}\longrightarrow \mathcal{W}_{3}-d\big(\sigma\mathcal{R}^{a}_{\,\,\,a}\big),
\label{eq:WeyltransWn=1}
\end{align}
where we used~$d\mathcal{R}^{a}_{\,\,\,a}=0$ [see eq.~\eqref{eq:0th_bianchi_trQ}].
Recasting the zeroth Bianchi identity~\eqref{eq:0th_bianchi_id} 
in the form~$D\mathcal{Q}^{a}_{\,\,\,b}
=\mathcal{R}^{a}_{\,\,\,b}+\mathcal{R}_{b}^{\,\,\,a}
+\mathcal{Q}^{ac}\wedge\mathcal{Q}_{cb}$ and using the second Bianchi 
identity in~\eqref{eq:1st2dn_bianchi_id},
we find
\begin{align}
d\mathcal{W}_{3}&={1\over 2}D\mathcal{Q}^{a}_{\,\,\,b}\wedge \mathcal{R}^{b}_{\,\,\,a} \nonumber \\[0.2cm]
&=\mathcal{R}_{(ab)}\wedge\mathcal{R}^{(ab)}
+{1\over 2}\mathcal{Q}^{ac}\wedge\mathcal{Q}_{cb}\wedge\mathcal{R}^{b}_{\,\,\,a}
\\[0.2cm]
&={1\over 4}D\mathcal{Q}_{ab}\wedge D\mathcal{Q}^{ab}
-{1\over 2}\mathcal{Q}^{ca}\wedge\mathcal{R}_{ab}\wedge\mathcal{Q}_{\,\,\,c}^{b}.
\nonumber
\end{align}
In fact, the symmetric part of the curvature two-form does not contribute to the second term in 
the last expression, so we 
define the invariant
\begin{align}
I^{Q}_{4}&\equiv {1\over 4}D\mathcal{Q}_{ab}\wedge D\mathcal{Q}^{ab}
-{1\over 2}\mathcal{Q}^{ca}\wedge\mathcal{R}_{[ab]}\wedge\mathcal{Q}_{\,\,\,c}^{b}.
\label{eq:IQ}
\end{align}
This quantity
can be regarded as a kind of nonmetricity counterpart to the
Nieh-Yan invariant and, by construction, is Weyl invariant\footnote{Unlike 
the torsional Nieh-Yan invariant, that
does change under Weyl transformations since the torsional Chern-Simons 
form~$\mathcal{H}=\theta_{a}\wedge T^{a}$ has Weyl
weight~$w=2$, i.e.,~$\mathcal{H}\rightarrow e^{2\sigma}\mathcal{H}$ (see
Appendix~\ref{sec:appendix}).}
[see eq.~\eqref{eq:WeyltransWn=1}].
In fact, it can be expressed as
\begin{align}
I_{4}^{Q}&=8\pi^{2}\Big[p_{1}(\widetilde{\mathcal{R}})-p_{1}(\mathcal{R})\Big],
\end{align}
where~$p_{1}(\mathcal{R})$ and~$p_{1}(\widetilde{\mathcal{R}})$ are, 
respectively, the first Pontrjagin classes of the
full connection~$\omega^{a}_{\,\,\,b}$ and the 
metric-compatible, Weyl invariant 
connection~$\widetilde{\omega}^{a}_{\,\,\,b}$ introduced above [see also 
eq.~\eqref{eq:omega_tilde_def}].

Implementing the decomposition~\eqref{eq:Q_Qbar_def}, eq.~\eqref{eq:IQ} is recast as
\begin{align}
I_{4}^{Q}&={1\over 4}\mathscr{D}\notrQ_{ab}\wedge \mathscr{D}\notrQ^{ab}
-{1\over 2}\notrQ^{ca}\wedge\notrR_{[ab]}\wedge\notrQ^{b}_{\,\,\,c}+d\mathcal{Q}\wedge d\mathcal{Q},
\label{eq:I4Q_decomp}
\end{align} 
where~$\mathscr{D}\notrQ_{ab}\equiv D\notrQ_{ab}+\mathcal{Q}\wedge\notrQ_{ab}$
is the Weyl covariant derivative defined in eq.~\eqref{eq:cov_der_def_new_param_Weyltrans},
and we used~$\eta^{ab}\mathscr{D}\notrQ_{ab}=0$ and~$\notrR_{[ab]}=\mathcal{R}_{[ab]}$. 
From this expression, we see that~$I_{4}^{Q}$ includes the second (unnormalized) Chern character 
of the Weyl gauge field~\cite{Nakahara:2003nw,AG_VM2023}, while
\begin{align}
\mathcal{W}_{3}
&={1\over 2}\notrQ^{a}_{\,\,\,b}\wedge\notrR^{b}_{\,\,\,a}+\mathcal{Q}\wedge d\mathcal{Q},
\end{align}
contains its corresponding Chern-Simons form.
 
In~$2n+2$ dimensions, the
nonmetricity Chern-Simons form can be generalized by writing
\begin{align}
\mathcal{W}_{2n+1}&={1\over 2}\mathcal{Q}^{a}_{\,\,\,b_{1}}\wedge \mathcal{R}^{b_{1}}_{\,\,\,\,b_{2}}
\wedge\stackrel{(n)}{\ldots}\wedge\mathcal{R}^{b_{n}}_{\,\,\,\,\,a},
\end{align}
that also changes by an exact differential under Weyl transformations
\begin{align}
\mathcal{W}_{2n+1}\longrightarrow \,\,& \mathcal{W}_{2n+1}-d\sigma\wedge 
{\rm tr\,}\mathcal{R}^{n} \nonumber \\[0.2cm]
&=\mathcal{W}_{2n+1}-d\big(\sigma\,{\rm tr\,}\mathcal{R}^{n}\big),
\end{align}
where we applied~$d\,{\rm tr\,}\mathcal{R}^{n}=0$
and the notation~$\zeta^{n}\equiv \zeta\,\wedge\stackrel{(n)}{\ldots}\wedge\,\zeta$ has been used, 
with matrix products implicitly understood. Similarly to the~$n=1$ case, we define
the nonmetricity invariant~$(2n+2)$-form
\begin{align}
I^{Q}_{2n+2}&\equiv d\mathcal{W}_{2n+1},
\end{align}
where~$I^{Q}_{2n+2}$ is explicitly given by
\begin{align}
I^{Q}_{2n+2}={1\over 2}D\mathcal{Q}_{ab_{1}}\wedge \mathcal{R}^{b_{1}}_{\,\,\,\,b_{2}}
\wedge\stackrel{(n)}{\ldots}\wedge\mathcal{R}^{b_{n}a}
-{1\over 4}\mathcal{Q}_{db_{1}}\wedge \mathcal{R}^{b_{1}}_{\,\,\,\,b_{2}}
\wedge\stackrel{(n)}{\ldots}\wedge\mathcal{R}^{b_{n}}_{\,\,\,\,\,a}\wedge\mathcal{Q}^{ad}.
\end{align}
Implementing the decomposition~\eqref{eq:Q_Qbar_def} once again, we find
\begin{align}
I^{Q}_{2n+2}&=\sum_{k=1}^{n}{1\over 2^{n-k+1}}{n\choose k}\Big(\mathscr{D}\notrQ^{ab_{1}}
\wedge\notrR_{b_{1}b_{2}}
\wedge\stackrel{(k)}{\ldots}\wedge\notrR^{b_{k}}_{\,\,\,\,\,a}
\nonumber \\[0.2cm]
&-\notrQ^{c}_{\,\,\,b_{1}}\wedge\notrR^{b_{1}}_{\,\,\,\,\,b_{2}}
\wedge\stackrel{(k)}{\ldots}\wedge\notrR^{b_{k}}_{\,\,\,\,\,a}\wedge\notrQ^{a}_{\,\,\,c}
\Big)\wedge (d\mathcal{Q})^{n-k}
\label{eq:In-dim_gen_decomp} \\[0.2cm]
&+\sum_{k=2}^{n}{1\over 2^{n-k+1}}{n\choose k}{\rm tr\,}(\notrR)^{k}\wedge 
(d\mathcal{Q})^{n-k+1}
+{n+1\over 2^{n}}(d\mathcal{Q})^{n+1}.
\nonumber
\end{align}
Note that the first term in the last line, 
coupling the Weyl field strength to the
background shear curvature, only exists for~$n\geq 2$ and it is therefore absent from
eq.~\eqref{eq:I4Q_decomp}, while the last contribution is the unnormalized $(n+1)$-th Chern 
character of the Weyl gauge field.

\section{Descent analysis}
\label{sec:anomaly_pol}

To apply our results to the microscopic couplings proposed 
in refs.~\cite{Rigouzzo:2023sbb,Palumbo:2024vqv} we need to consider 
the ``pure Weyl'' case\footnote{As discussed in sec.~\ref{sec:closing}, 
the constitutive relation for the chiral coupling of ref.~\cite{Rigouzzo:2023sbb}
can be obtained from the one for the model proposed in~\cite{Palumbo:2024vqv}
by an appropriate replacement of the Weyl gauge field.},~$\notrQ_{ab}=0$. Nevertheless, 
for the time being
we are going to be more general and use the general
invariant~\eqref{eq:IQ} to write the anomaly polynomial
\begin{align}
\mathcal{P}_{6}(\mathcal{F}_{A},I^{Q}_{4})=c_{W}\mathcal{F}_{A}\wedge I^{Q}_{4},
\label{eq:anomaly_polynomial_P6}
\end{align}
where~$\mathcal{F}_{A}=d\mathcal{A}$ is the field strength of the external gauge field coupling to the
axial-vector fermion current
and~$c_{W}$ is a 
pure number to be determined from the details of the coupling. 
This anomaly polynomial is invariant under both axial-vector gauge
transformations~$\delta_{\gamma}\mathcal{A}=d\gamma$, with~$\gamma$ an arbitrary
zero-form, and Weyl transformations, that leave the gauge field~$\mathcal{A}$ unchanged
[see the discussion after eq.~\eqref{eq:IQ}]. 
The associated Chern-Simons five-form
\begin{align}
\omega^{0}_{5}(\mathcal{A},I^{Q}_{4})=c_{W}\mathcal{A}\wedge I^{Q}_{4},
\label{eq:CS_5form_full}
\end{align} 
leads to the following expression of the consistent anomaly for the axial-vector current coupling to~$\mathcal{A}$
\begin{align}
\delta_{\gamma}\omega^{0}_{5}&= d\omega^{1}_{4}
\hspace*{1cm} \Longrightarrow \hspace*{1cm} \omega_{4}^{1}=c_{W}\gamma I^{Q}_{4}.
\label{eq:consistent_anomaly_gen}
\end{align}
In writing the Chern-Simons form~\eqref{eq:CS_5form_full}, we have
chosen to preserve its invariance under Weyl transformations, at the expense of
axial-vector gauge invariance. In fact, we can shift 
from one invariance to the other by adding to~$\omega^{0}_{5}$ an exact differential
\begin{align}
\omega^{0}_{5}(a)\equiv
\omega^{0}_{5}-{ac_{W}\over 2}d\big(\mathcal{A}\wedge \mathcal{W}_{3}\big),
\label{eq:CS(a)_form}
\end{align}
with~$a$ a real parameter. Indeed,
for~$a=-2$ the corresponding Chern-Simons form~$\omega^{0}_{5}(-2)=c_{W}\mathcal{F}_{A}\wedge\mathcal{W}_{3}$ 
is invariant under
axial-vector gauge transformations but not under Weyl rescalings~\eqref{eq:WeyltransWn=1}.
The associated consistent anomaly for general~$a$ is given by
\begin{align}
\omega^{1}_{4}(a)={(2+a)c_{W}\over 2}I^{Q}_{4},
\end{align}
and vanishes when~$a=-2$. Defining the Chern-Simons nonlocal effective action
\begin{align}
\Gamma(a)_{\rm CS}=\int_{\mathcal{M}_{5}}\omega^{0}_{5}(a),
\end{align}
where~$\mathcal{M}_{5}$ is an Euclidean five-dimensional manifold
whose boundary~$\partial\mathcal{M}_{5}$ is identified with the compatified four-dimensional spacetime, the 
integrated consistent
anomaly is obtained from its gauge variation
\begin{align}
\delta_{\gamma}\Gamma(a)_{\rm CS}&=\int_{\partial\mathcal{M}_{5}}\omega^{1}_{4}(a)
\nonumber \\[0.2cm]
&={(2+a)c_{W}\over 2}\int_{\partial\mathcal{M}_{5}}\gamma I^{Q}_{4}  \\[0.2cm]
&\equiv -\int_{\partial\mathcal{M}_{5}}\gamma d\langle\star J_{5}\rangle_{\rm cons},
\nonumber
\end{align}
where~$\star$ indicates the four-dimensional Hodge dual.

The dependence of the consistent anomaly on the local counterterm
parameter~$a$ is not shared by the covariant form of the anomaly.
To find the covariant current, 
we need to compute first the Bardeen-Zumino (BZ) term~\cite{Bardeen:1984pm}, 
given by (see, e.g.,~\cite{Jensen:2013kka,Manes:2018llx})
\begin{align}
\langle\star J_{5}\rangle_{\rm BZ}&\equiv -{\delta\over \delta\mathcal{F}_{A}} \Gamma(a)_{\rm CS}
\nonumber \\[0.2cm]
&={ac_{W}\over 2}\mathcal{W}_{3}.
\label{eq:BZ_current_general}
\end{align}
Adding this expression to the one of the consistent 
current,~$\langle\star J_{5}\rangle_{\rm cons}=-{c_{W}(a+2)\over 2}\mathcal{W}_{3}$, obtained
from the boundary variation of the effective action under~$\mathcal{A}\rightarrow
\mathcal{A}+\delta\mathcal{A}$,
we get the covariant current 
\begin{align}
\langle\star J_{5}\rangle_{\rm cov}&=-c_{W}\mathcal{W}_{3}.
\end{align}
This is independent
of~$a$, as it is also the case for the covariant anomaly
\begin{align}
d\langle\star J_{5}\rangle_{\rm cov}=-c_{W}I^{Q}_{4}.
\label{eq:cov_anomaly_gen}
\end{align}
Since for applications to chiral transport we are 
interested
in the constitutive relation of the
covariant axial-vector current~\cite{Manes:2019fyw}, we set~$a=0$ from now on.

\section{The chiral current constitutive relation}
\label{sec:constitutive_relations}

\paragraph{General analysis.}
We are now ready to address the problem of nonmetricity-induced transport of chiral charge
in a fermion fluid at thermal equilibrium.
The equilibrium partition function will be computed from the Chern-Simons five-form~\eqref{eq:CS_5form_full} 
by using the transgression
methods presented, e.g., in refs.~\cite{Jensen:2013kka,Manes:2018llx,Valle:2021nfv}, where the reader can 
find full details.
We will eventually particularize our expressions to the ``pure Weyl'' case,~$\notrQ_{ab}=0$,
in order to apply them to the models proposed in refs.~\cite{Palumbo:2024vqv,Rigouzzo:2023sbb}.

We consider a one-parameter family of connections ($0\leq t\leq 1$)
\begin{align}
\mathcal{A}_{t}&=\widehat{\mathcal{A}}-t\mu_{5}u, \nonumber \\[0.2cm]
(\omega_{t})^{a}_{\,\,\,b}&=\widehat{\omega}^{a}_{\,\,\,b}-t\mu^{a}_{\,\,\,b}u, 
\label{eq:A_tomega_t}
\end{align}
where~$u\equiv u_{\mu}dx^{\mu}$ is the fluid's four-velocity one-form
and~$\mu_{5}$ and~$\mu_{ab}$ are the chiral and spin chemical potentials, respectively.
The associated field strength, curvature, and nonmetricity have the following expressions
\begin{align}
\mathcal{F}_{t}&=d\widehat{\mathcal{A}}
-2t\mu_{5}\omega+tu\big(d+\mathfrak{a}\big)\mu_{5}\omega, \nonumber \\[0.2cm]
(\mathcal{R}_{t})^{a}_{\,\,\,b}
&= \widehat{\mathcal{R}}^{a}_{\,\,\,b}
-2t\mu^{a}_{\,\,\,b}\omega
+tu\wedge \big(\widehat{D}+\mathfrak{a}\big)\mu^{a}_{\,\,\,b},
\\[0.2cm]
(\mathcal{Q}_{t})_{ab}&=\widehat{\mathcal{Q}}_{ab}-t(\mu_{Q})_{ab}u,
\nonumber
\end{align}
where
\begin{align}
\widehat{\mathcal{R}}^{a}_{\,\,\,b}&\equiv
d\widehat{\omega}^{a}_{\,\,\,b}
+\widehat{\omega}^{a}_{\,\,\,c}\wedge\widehat{\omega}^{c}_{\,\,\,b}, \nonumber \\[0.2cm]
\widehat{D}\mu^{a}_{\,\,\,b}&\equiv d\mu^{a}_{\,\,\,b}
+\widehat{\omega}^{a}_{\,\,\,c}\mu^{c}_{\,\,\,b}-\mu^{a}_{\,\,\,c}\widehat{\omega}^{c}_{\,\,\,b},
\label{eq:Rhat_definition_first}
\end{align}
while~$\mathfrak{a}\equiv u\rfloor du$ and~$\omega\equiv {1\over 2}(du+u\wedge\mathfrak{a})$ 
are, respectively, the fluid's acceleration and
vorticity.
The definition of the nonmetricity one-form~\eqref{eq:nonmetricity_def} leads to the relations
\begin{align}
\widehat{\mathcal{Q}}_{ab}&=2\widehat{\omega}_{(ab)}, \nonumber \\[0.2cm]
(\mu_{Q})_{ab}&=2\mu_{(ab)}.
\label{eq:Qhatvssymomega_def}
\end{align}

To evaluate the Chern-Simons form, 
we apply the Ma\~nes-Stora-Zumino generalized transgression formula~\cite{Manes:1985df} 
(see Appendix A of ref.~\cite{Manes:2018llx} for 
a quick overview)
\begin{align}
\int_{\partial T}{\ell_{t}^{p}\over p!}\mathscr{Q}=\int_{T}{\ell_{t}^{p+1}\over (p+1)!}d\mathscr{Q}+(-1)^{p+q}
d\int_{T}{\ell_{t}^{p+1}\over (p+1)!}\mathscr{Q},
\label{eq:generalized_transgression}
\end{align}
where~$\mathscr{Q}$ is a polynomial depending on the family of connections and their
curvatures, and~$\ell_{t}$ is an even operator
replacing the exterior differential~$d$ by the odd operator
\begin{align}
d_{t}=dt{d\over dt}.
\end{align}
In addition,~$p+1$ is the number of parameters in the family of connections and~$q$ is the order
of the polynomial~$\mathscr{Q}$ in the operator~$d_{t}$.
We use eq.~\eqref{eq:generalized_transgression} with
\begin{align}
\mathscr{Q}&=\omega_{0}^{5}(\mathcal{A}_{t},I^{Q}_{4,t}),\nonumber \\[0.2cm]
d\mathscr{Q}&=\mathcal{P}_{6}(\mathcal{A}_{t},I^{Q}_{4,t}),
\end{align}
where~$I^{Q}_{4,t}=d\mathcal{W}_{3,t}$ and~$\mathcal{W}_{3,t}={1\over 2}(\mathcal{Q}^{a}_{\,\,\,b})_{t}
\wedge(\mathcal{R}^{b}_{\,\,\,a})_{t}$.
Considering that in our case~$T=[0,1]$ and~$p=q=0$, the transgression 
five-form can be written as
\begin{align}
\omega_{5}^{0}(\mathcal{A}_{t=1},I^{Q}_{4,t=1})
-\omega_{5}^{0}(\mathcal{A}_{t=0},I^{Q}_{4,t=0})
&=\int_{0}^{1}\ell_{t}\mathcal{P}_{6}(\mathcal{A}_{t},I^{Q}_{4,t})
+d\int_{0}^{1}\ell_{t}\omega_{5}^{0}(\mathcal{A}_{t},I^{Q}_{4,t}).
\label{eq:transgression_def}
\end{align}
At equilibrium we have a general stationary background
whose line element takes the form~$ds^{2}=u\otimes u+g_{ij}(\mathbf{x})dx^{i}\otimes dx^{j}$ 
and all hatted quantities become purely magnetic, i.e., transverse to~$u$, a fact 
indicated by boldface fonts. We thus write
\begin{align}
\mathcal{A}_{t=1}&=\boldsymbol{A}-u\mu_{5}, \nonumber \\[0.2cm]
(\omega_{t=1})^{a}_{\,\,\,b}&=\boldsymbol{\omega}^{a}_{\,\,\,b}-\mu^{a}_{\,\,\,b}u, \nonumber \\[0.2cm]
\mathcal{F}_{t=1}&=d\boldsymbol{A}
-2\mu_{5}\omega+u\big(d+\mathfrak{a}\big)\mu_{5}  \nonumber \\[0.2cm]
&\equiv \boldsymbol{B}_{A}+u\wedge E_{A},\nonumber \\[0.2cm]
(\mathcal{R}_{t=1})^{a}_{\,\,\,b}&=
d\boldsymbol{\omega}^{a}_{\,\,\,b}+\boldsymbol{\omega}^{a}_{\,\,\,c}\wedge
\boldsymbol{\omega}^{c}_{\,\,\,b}
-2\mu^{a}_{\,\,\,b}\omega+u\wedge 
\big(\boldsymbol{D}+\mathfrak{a}\big)\mu^{a}_{\,\,\,b} \label{eq:equ_identities_var} \\[0.2cm]
&\equiv \boldsymbol{B}^{a}_{\,\,\,b}+u\wedge E^{a}_{\,\,\,b},
\nonumber \\[0.2cm]
(\mathcal{Q}_{t=1})_{ab}&=\boldsymbol{Q}_{ab}-(\mu_{Q})_{ab}u, \nonumber \\[0.2cm]
D(\mathcal{Q}_{t=1})_{ab}&=\boldsymbol{D}\boldsymbol{Q}_{ab}-2(\mu_{Q})_{ab}\omega
+u\wedge\Big[\big(\boldsymbol{D}+\mathfrak{a}\big)(\mu_{Q})_{ab}+\boldsymbol{Q}_{cb}\mu^{c}_{\,\,\,a}
+\boldsymbol{Q}_{ac}\mu^{c}_{\,\,\,b}\Big] \nonumber \\[0.2cm]
&= 2\boldsymbol{B}_{(ab)}+2u\wedge E_{(ab)},
\nonumber
\end{align}
where all quantities are time independent 
and~$\boldsymbol{D}$ is the exterior covariant differential associated to the magnetic part of the
connection,~$\boldsymbol{\omega}^{a}_{\,\,\,b}$.
Moreover, since~$\omega_{5}^{0}(\mathcal{A}_{t=0},I^{Q}_{4,t=0})$ is a maximal form with no
component along~$u$, it vanishes. The equilibrium partition function is then
obtained by integrating eq.~\eqref{eq:transgression_def}
over the five-dimensional bulk Euclidean space~$\mathcal{M}_{5}$
\begin{align}
W_{\rm eq}&=\int_{\mathcal{M}_{5}}\omega_{5}^{0}(\mathcal{A}_{t=1},I^{Q}_{4,t=1}) \nonumber \\[0.2cm]
&=\int_{\mathcal{M}_{5}}\int_{0}^{1}\ell_{t}\mathcal{P}_{6}(\mathcal{A}_{t},I^{Q}_{4,t})
+\int_{\partial\mathcal{M}_{5}}\int_{0}^{1}\ell_{t}\omega_{5}^{0}(\mathcal{A}_{t},I^{Q}_{4,t}) \\[0.2cm]
&\equiv W_{\rm bulk}+W_{\rm bdy}.
\nonumber
\end{align}
A rather lengthy computation of the terms in the second line renders
the following expressions for the bulk and boundary components of the partition function
\begin{align}
W_{\rm bulk}&=
-{c_{W}\over 2}\int_{\mathcal{M}_{5}}u\wedge\Bigg\{\mu_{5}
\bigg[2\notrB^{(ab)}\wedge\notrB_{(ab)} 
+4(\notrmu_{Q})^{ab}\notrB_{(ab)}\wedge\omega
\nonumber \\[0.2cm]
&-\notrQbf^{ca}\wedge\boldsymbol{B}_{[ab]}\wedge\notrQbf^{b}_{\,\,\,c}
-2\mu_{[ab]}\,\notrQbf^{ac}\wedge\,\notrQbf_{c}^{\,\,\,b}\wedge\omega
+2(\notrmu)^{ab}(\notrmu)_{ab}\,\omega\wedge\omega \Big] \nonumber \\[0.2cm]
&+\Big[2(\notrmu_{Q})^{ab}\notrB_{(ab)}
-\mu_{[ab]}\,\notrQbf^{ac}\wedge\notrQbf_{c}^{\,\,\,b}
+(\notrmu_{Q})^{ab}(\notrmu_{Q})_{ab}\,\omega\Big]\wedge\boldsymbol{B}_{A}
\Bigg\} \label{eq:W_bulk_decomp} \\[0.2cm]
&-c_{W}\int_{\mathcal{M}_{5}}u\wedge\bigg[ 
\mu_{5}\Big(\boldsymbol{B}_{Q}\wedge\boldsymbol{B}_{Q} 
+4\mu_{Q}\boldsymbol{B}_{Q}\wedge\omega+4\mu_{Q}^{2}\,\omega\wedge\omega\Big)
+2\mu_{Q}\big(\boldsymbol{B}_{Q}+\mu_{Q}\omega\big)\wedge\boldsymbol{B}_{A}
\bigg],
\nonumber
\end{align}
and
\begin{align}
W_{\rm bdy}
&={c_{W}\over 2}\int_{\partial\mathcal{M}_{5}}
\,u\wedge\Big[2(\notrmu_{Q})^{ab}\notrB_{(ab)}
-\mu_{[ab]}\notrQbf^{ac}\wedge\notrQbf_{c}^{\,\,\,b}
+(\notrmu_{Q})^{ab}(\notrmu_{Q})_{ab}\omega 
\Big]\wedge\boldsymbol{A} \nonumber \\[0.2cm]
&+2c_{W}\int_{\partial\mathcal{M}_{5}}
\,u\wedge\big(\mu_{Q}\boldsymbol{B}_{Q}
+\mu_{Q}^{2}\omega \big)\wedge\boldsymbol{A}.
\label{eq:W_bdy_decomp}
\end{align}
In these expressions we have separated the trace and traceless parts of all quantities,
and introduced 
the chemical potential associated to the Weyl gauge field
\begin{align}
\mathcal{Q}=\boldsymbol{Q}-\mu_{Q}u, 
\end{align}
given by
\begin{align}
\mu_{Q}\equiv {1\over 4}(\mu_{Q})^{a}_{\,\,\,a},
\end{align}
together with 
\begin{align}
\boldsymbol{B}_{Q}&\equiv {1\over 2}\boldsymbol{B}^{a}_{\,\,\,a}
=d\boldsymbol{Q}-2\mu_{Q}\omega.
\label{eq:B_Q_definition}
\end{align}
The covariant axial-vector current is obtained from the boundary contribution of
the functional variation of bulk partition function 
with respect to~$\boldsymbol{A}$~\cite{Jensen:2013kka,Manes:2018llx}, with the result
\begin{align}
\langle \star J_{5}\rangle_{\rm cov}
&={c_{W}\over 2}u\wedge\Big[2(\notrmu_{Q})^{ab}\notrB_{(ab)}
-\mu_{[ab]}\notrQbf^{ac}\wedge\notrQbf_{c}^{\,\,\,b}
+(\notrmu_{Q})^{ab}(\notrmu_{Q})_{ab}\,\omega\Big] \nonumber \\[0.2cm]
&+2c_{W}u\wedge\Big(\mu_{Q}\boldsymbol{B}_{Q}
+\mu_{Q}^{2}\omega\Big).
\label{eq:const_relJ_decomp}
\end{align}
A computation of the consistent current from the boundary partition function~\eqref{eq:W_bdy_decomp}
shows that~$\langle\star J_{5}\rangle_{\rm cons}=\langle\star J_{5}\rangle_{\rm cov}$, 
which is consistent with the expression of
the BZ current given in eq.~\eqref{eq:BZ_current_general} for~$a=0$.

The covariant anomaly is then evaluated to give
\begin{align}
d\langle \star J_{5}\rangle_{\rm cov}
&=-{c_{W}\over 2}u\wedge\Big[
4\notrE^{(ab)}\wedge\notrB_{(ab)}
+2(\notrmu_{Q})_{ac}\,\notrQbf^{a}_{\,\,\,b}\wedge\notrB^{[cb]}
+\notrQbf^{ac}\wedge \notrE_{[ab]}\wedge\notrQbf_{\,\,\,c}^{b}
\Big] \nonumber \\[0.2cm]
&-2c_{W}u\wedge
E_{Q}\wedge\boldsymbol{B}_{Q},
\end{align}
where
\begin{align}
E_{Q}&\equiv {1\over 2}E^{a}_{\,\,\,a}=(d+\mathfrak{a})\mu_{Q}.
\end{align}
A straightforward calculation shows that this is in full agreement 
with the general result given in eq.~\eqref{eq:cov_anomaly_gen}.

The covariant chiral current whose constitutive relation we evaluated
in~\eqref{eq:const_relJ_decomp} is, up to a global numerical normalization, the dilation piece of the 
(covariant)
hypermomentum current~\cite{Hehl:1976_1,Hehl:1976_2,Hehl:1976_3,Gronwald:1991st,Obukhov:1993pt}
\begin{align}
\langle\Delta\rangle_{\rm cov}&=2\langle \star J_{5}\rangle_{\rm cov}.
\end{align}
Its shear part, on the other hand, is obtained from the boundary contribution of the variation 
of~$W_{\rm bulk}$ with respect to~$\notrQbf_{ab}$, giving
\begin{align}
\langle\,\notrDelta^{ab}\rangle_{\rm cov}
\equiv \left.2{\delta W_{\rm bulk}\over \delta\notrQbf_{ab}}\right|_{\rm bdy}
=2c_{W}u\wedge\Big\{2\mu_{5}
\big[\notrB^{(ab)} 
+(\notrmu_{Q})^{ab}\omega\big]
+(\notrmu_{Q})^{ab}\boldsymbol{B}_{A}
\Big\},
\label{eq:shear_hypermomentum}
\end{align}
whereas the calculation for the spin current piece renders
\begin{align}
\langle\sigma^{ab}\rangle_{\rm cov}
\equiv \left.{\delta W_{\rm bulk}\over \delta\boldsymbol{\omega}_{[ab]}}\right|_{\rm bdy}
&=-{c_{W}\over 2}u\wedge\notrQbf^{ac}\wedge\notrQbf^{\,\,\,b}_{c}.
\label{eq:spin_hypermomentum}
\end{align}
Using the transformations of~$\notrQ^{a}_{\,\,\,b}$ and~$\mathcal{Q}$ shown in
eq.~\eqref{eq:notrQransf_Q_trans_general_app}, we find that both the bulk and
boundary equilibrium partition functions,~\eqref{eq:W_bulk_decomp} 
and~\eqref{eq:W_bdy_decomp}, as well as the constitutive relation of the 
chiral current~\eqref{eq:const_relJ_decomp} and the hypermomentum current 
components~\eqref{eq:shear_hypermomentum} and~\eqref{eq:spin_hypermomentum}, 
are invariant under time-independent
Weyl transformation~($\Lambda=1$,~$\Xi=0$).

\paragraph{The case of Weyl-type nonmetricity.}
Our discussion so far has remained rather formal. To study the consequences for 
chiral transport of the coupling of fermionic matter to nonmetricity put forward
in~\cite{Palumbo:2024vqv}, 
however, we need to set the shear nonmetricity component to zero,~$\notrQbf_{ab}
=(\notrmu_{Q})_{ab}=0$. 
Recasting the zeroth Bianchi identity~\eqref{eq:Bianchi_id_overlineQ1} as
\begin{align}
2\notrB_{(ab)}&=\pmb{\mathscr{D}}\notrQbf_{ab}-2(\notrmu_{Q})_{ab}\omega, \nonumber \\[0.2cm]
2\notrE_{(ab)}&=(\pmb{\mathscr{D}}+\mathfrak{a})(\notrmu_{Q})_{ab}
+\notrmu_{ca}\notrQbf^{c}_{\,\,\,b}
+\notrmu_{cb}\notrQbf^{c}_{\,\,\,a},
\end{align}
with~$\pmb{\mathscr{D}}$ the Weyl covariant differential~\eqref{eq:cov_der_def_new_param_Weyltrans}
built from the magnetic part of the spin
connection~$\boldsymbol{\omega}^{a}_{\,\,\,b}$, 
we find that~$\notrE_{(ab)}=\notrB_{(ab)}=0$. The bulk and boundary equilibrium partition functions
then take the form
\begin{align}
W_{\rm bulk}&=
-c_{W}\int_{\mathcal{M}_{5}}u\wedge\bigg[ 
\mu_{5}\Big(\boldsymbol{B}_{Q}\wedge\boldsymbol{B}_{Q} 
+4\mu_{Q}\boldsymbol{B}_{Q}\wedge\omega+4\mu_{Q}^{2}\,\omega\wedge\omega\Big)
+2\mu_{Q}\big(\boldsymbol{B}_{Q}+\mu_{Q}\omega\big)\wedge\boldsymbol{B}_{A}
\bigg], \nonumber \\[0.2cm]
W_{\rm bdy}&=2c_{W}\int_{\partial\mathcal{M}_{5}}
\,u\wedge\big(\mu_{Q}\boldsymbol{B}_{Q}
+\mu_{Q}^{2}\omega \big)\wedge\boldsymbol{A},
\label{eq:W_Q_case1}
\end{align} 
while the constitutive relation for the covariant axial-vector current, obtained by functional
differentiation of the bulk action with respect to~$\boldsymbol{B}_{A}$,
reads
\begin{align}
\langle\star J_{5}\rangle_{\rm cov}&=2c_{W}u\wedge\big(\mu_{Q}\boldsymbol{B}_{Q}+\mu_{Q}^{2}\omega\big).
\label{eq:constitutive_rel_J5}
\end{align}
Furthermore, a look at eqs.~\eqref{eq:shear_hypermomentum} and~\eqref{eq:spin_hypermomentum} shows 
that~$\langle\Delta\rangle_{\rm cov}=2\langle\star J_{5}\rangle_{\rm cov}$ 
is the only piece of the hypermomentum current surviving the ``pure Weyl'' limit.

The constitutive relation~\eqref{eq:constitutive_rel_J5} 
reveals the existence of two sources of chiral charge transport induced by
nonmetricity, respectively driven by the Weyl magnetic field
and the fluid's vorticity. The corresponding transport coefficients of 
these nonmetricity magnetic and chiral separation effects
depend on the nonmetricity chemical potential, determined by 
the time component of the Weyl gauge field. 
Finally, comparing the expression of the axial anomaly 
found in ref.~\cite{Palumbo:2024vqv} with the one given in eq.~\eqref{eq:cov_anomaly_gen} 
for~$\notrQ_{ab}=0$, shows
the constant~$c_{W}$ to be given by
\begin{align}
c_{W}=-{e^{2}\over 4\pi^{2}},
\end{align}
with~$e$ the electron charge.

\section{Discussion}
\label{sec:closing}

We have investigated the consequences of background nonmetricity for the transport 
of chiral charge in fermionic fluids at equilibrium.
Our main result is the existence of magnetic and vortical nonmetricity-driven 
chiral separation effects that, in view of the results of refs.~\cite{Palumbo:2024vqv,Pires:2022fdf}, 
might be of physical relevance 
for the context of Weyl semimetals with point defects. 
Both effects
disappear for static geometries with~$\mathcal{Q}_{0}=0$ (see the related
discussion in ref.~\cite{Palumbo:2024vqv}). Our analysis here, however, 
considers stationary backgrounds in which~$\mathcal{Q}_{0}$, and hence~$\mu_{Q}$,
is generically nonzero.

In the previous section, we presented the analysis of the model
proposed in~\cite{Palumbo:2024vqv}, where the 
axial-vector fermionic current couples the Weyl gauge field. The
nonminimal chiral coupling of 
ref.~\cite{Rigouzzo:2023sbb}, on the other hand, also involves the
second gauge field~$\widehat{\mathcal{Q}}_{a}$, defined in eq.~\eqref{eq:second_nm_gauge_field}, 
as well as the edge torsion~$\mathcal{T}_{a}$. 
A comparison between the actions~\eqref{eq:action_palumbo} 
and~\eqref{eq:action_nonminimal_rigouzzo_zell} shows that the 
results for
the latter model can be found from our analysis
by replacing~$\mathcal{Q}\rightarrow {1\over 2}\beta\widetilde{\mathcal{T}}_{a}\theta^{a}$
[see eq.~\eqref{eq:Bagaugefield}]. In particular, the corresponding constitutive relation for the
chiral current is obtained from~\eqref{eq:constitutive_rel_J5} by substituting
\begin{align}
\mu_{Q}&\longrightarrow {\beta\over 4}\big(2\mu_{\mathcal{T}}+\mu_{\mathcal{\widehat{Q}}}-4\mu_{Q}\big),
\nonumber \\[0.2cm]
\boldsymbol{B}_{Q}&\longrightarrow
{\beta\over 4}\big(2\boldsymbol{B}_{\mathcal{T}}+\boldsymbol{B}_{\widehat{Q}}
-4\boldsymbol{B}_{Q}\big),
\end{align}  
where the chemical potentials~$\mu_{\mathcal{T}}$ and~$\mu_{\widehat{Q}}$ are defined by the 
electric-magnetic decompositions of the screw torsion~$\mathcal{T}\equiv \mathcal{T}_{a}\theta^{a}$
and the second nonmetricity gauge field~$\widehat{\mathcal{Q}}$
\begin{align}
\mathcal{T}&\equiv \boldsymbol{\mathcal{T}}-\mu_{\mathcal{T}}u, \nonumber \\[0.2cm]
\widehat{\mathcal{Q}}&\equiv \widehat{\boldsymbol{Q}}-\mu_{\widehat{Q}}u,
\end{align}
while~$\boldsymbol{B}_{\mathcal{T}}$ and~$\boldsymbol{B}_{\widehat{Q}}$ are the 
magnetic components of their corresponding field strengths [cf. eq.~\eqref{eq:B_Q_definition}].
Interestingly, in this case the two chiral separation effects survive 
in the limit of vanishing nonmetricity, when they are driven by 
the background's edge torsion~$\mathcal{T}$ alone. This is in addition to chiral transport effects
resulting from the minimal chiral coupling of microscopic fermions
to the screw torsion gauge field~$\mathcal{S}_{a}$, that were investigated in~\cite{Valle:2021nfv}.

Together with the analysis of two specific fermionic couplings to 
nonmetricity~\cite{Palumbo:2024vqv,Rigouzzo:2023sbb}
 proposed in the 
literature, 
we have also presented a general descent formalism based on a nonmetricity Chern-Simons form
leading to the four-form nonmetricity invariant given
in eq.~\eqref{eq:IQ}. Besides its purely geometrical 
interest, 
the natural anomaly polynomial constructed from this invariant
could be instrumental in analyzing the physics of future
theories of fermionic matter chirally coupled to more general metric-affine backgrounds, 
involving this time the full shear nonmetricity tensor.
The framework recently put forward in~\cite{Wheeler:2026xfx} might provide the appropriate basis 
to search for such generalizations.

\section*{Acknowledgments} 

Illuminating discussions with Friedrich W. Hehl and Manuel Valle are gratefully acknowledged. 
This work has been supported by the Spanish Science Ministry Grant No. 
PID2024-160856NB-I00 (MCIU/AEI/FEDER, EU) and the Basque Government Grant No. IT1628-22.

\appendix

\section{Some identities in metric-affine geometry}
\label{sec:nonmetricity_generalities}

In what follows, we summarize basic identities in
metric-affine geometry 
relevant for our discussion (full details can be found, e.g., 
in~\cite{Hehl:1994ue,Ponomarev2017,JimenezCano:2021rlu}). Applying the 
general expression of the covariant 
differential~\eqref{eq:covariant_dif1} to
the definition of the nonmetricity tensor in 
eq.~\eqref{eq:nonmetricity_def}, we are led to the so-called 
zeroth Bianchi identity
\begin{align}
D\mathcal{Q}_{ab}=2\mathcal{R}_{(ab)},
\label{eq:0th_bianchi_id}
\end{align}
where, to compute the right-hand side, we used
\begin{align}
D^{2}\mathfrak{T}^{a_{1}\ldots a_{p}}_{\hspace*{0.8cm}b_{1}\ldots b_{q}}&=
\mathcal{R}^{a_{1}}_{\,\,\,\,c}\wedge
\mathfrak{T}^{c\ldots a_{p}}_{\hspace*{0.8cm}b_{1}\ldots b_{q}}+\ldots
+\mathcal{R}^{a_{p}}_{\,\,\,\,c}\wedge
\mathfrak{T}^{a_{1}\ldots c}_{\hspace*{0.8cm}b_{1}\ldots b_{q}}
 \nonumber \\[0.2cm]
&-\mathfrak{T}^{a_{1}\ldots a_{p}}_{\hspace*{0.8cm}c\ldots b_{q}}
\wedge \mathcal{R}^{c}_{\,\,\,b_{1}}-\ldots
-\mathfrak{T}^{a_{1}\ldots a_{p}}_{\hspace*{0.8cm}b_{1}\ldots c}\wedge 
\mathcal{R}^{c}_{\,\,\,\,b_{q}}.
\end{align}
The first and second Bianchi identities, respectively involving torsion 
and curvature, retain their standard form 
\begin{align}
DT^{a}&=\mathcal{R}^{a}_{\,\,\,b}\wedge \theta^{b}, \nonumber \\[0.2cm]
D\mathcal{R}^{a}_{\,\,\,b}&=0,
\label{eq:1st2dn_bianchi_id}
\end{align}
where~$T^{a}\equiv D\theta^{a}$ is the torsion
two-form. It is important, however, to keep in mind that in the presence of
nonmetricity  
we should be wary when lowering and rising indices inside covariant 
derivatives, since
extra terms might arise. 
For example, from~$D\eta_{ab}=-\mathcal{Q}_{ab}$ we find~$D\delta^{a}_{b}=0$, 
while by lowering the contravariant index 
the second Bianchi identity in the second line of eq.~\eqref{eq:1st2dn_bianchi_id} we arrive   
at~$D\mathcal{R}_{ab}=-\mathcal{Q}_{ac}\wedge\mathcal{R}^{c}_{\,\,\,b}$.  

The zeroth Bianchi identity~\eqref{eq:0th_bianchi_id} can be recast in terms of the
trace and the shear components of~$\mathcal{Q}_{ab}$ 
introduced in eq.~\eqref{eq:Q_Qbar_def}.   
The Weyl one-form,~$\mathcal{Q}={1\over 4}\mathcal{Q}^{a}_{\,\,\,a}$, 
satisfies
\begin{align}
d\mathcal{Q}&={1\over 2}\mathcal{R}^{a}_{\,\,\,a},
\label{eq:0th_bianchi_trQ}
\end{align}
therefore playing the role of a gauge potential 
for the trace of the curvature two-form.
As for the shear nonmetricity tensor, its Bianchi identity reads
\begin{align}
D\notrQ_{ab}+\mathcal{Q}\wedge\notrQ_{ab}
=2\notrR_{(ab)},
\label{eq:Bianchi_id_overlineQ1}
\end{align}
where we have used the notation defined in~\eqref{eq:Rbar_def}.

The
spin connection components~$\omega^{a}_{\,\,\,b}\equiv \omega_{c\,\,\,b}^{\,\,\,a}\theta^{c}$
can be split 
as~\cite{Hehl:1994ue,Ponomarev2017,JimenezCano:2021rlu}
\begin{align}
\omega^{\,\,\,b}_{a\,\,\,c}&=\oomega^{\,\,\,b}_{a\,\,\,c}+K^{\,\,\,b}_{a\,\,\,c}+L^{\,\,\,b}_{a\,\,\,c},
\label{eq:omega=oomega+distorsion_tensor}
\end{align}
where~$\oomega^{b}_{\,\,\,c}=\oomega^{\,\,\,b}_{a\,\,\,c}\theta^{a}$ is 
the Levi-Civita spin connection, 
while~$K^{\,\,\,b}_{a\,\,\,c}$ and~$L^{\,\,\,b}_{a\,\,\,c}$ are the contorsion
and disformation tensors, respectively. 
In terms of the torsion and nonmetricity components
\begin{align}
T^{a}&\equiv {1\over 2}T_{bc}^{\,\,\,\,\,a}\theta^{b}\wedge \theta^{c}, \nonumber \\[0.2cm]
\mathcal{Q}^{a}_{\,\,\,b}&\equiv \mathcal{Q}_{c\,\,\,b}^{\,\,\,a}\theta^{c},
\end{align}
these two tensors are explicitly given by
\begin{align}
K_{c\,\,\,\,b}^{\,\,\,a}&={1\over 2}\big(T^{a}_{\,\,\,\,cb}+T^{a}_{\,\,\,\,bc}-T_{bc}^{\,\,\,\,a}\big), 
\nonumber \\[0.2cm]
L_{c\,\,\,\,b}^{\,\,\,a}&={1\over 2}\big(\mathcal{Q}_{c\,\,\,\,b}^{\,\,\,a}
+\mathcal{Q}_{b\,\,\,\,c}^{\,\,\,a}-\mathcal{Q}_{\,\,\,bc}^{a}\big).
\label{eq:contortion+disformation}
\end{align}
Taking into account that~$T_{bca}=T_{[bc]a}$
and~$\mathcal{Q}_{cab}=\mathcal{Q}_{c(ab)}$, we find
\begin{align}
K_{cab}&=K_{c[ab]}, \nonumber \\[0.2cm]
L_{c(ab)}&={1\over 2}\mathcal{Q}_{cab}, \\[0.2cm]
L_{c[ab]}&={1\over 2}\big(\mathcal{Q}_{abc}-\mathcal{Q}_{bca}\big).
\nonumber 
\end{align}
On the other hand, from 
the antisymmetry of the Levi-Civita connection,~$\oomega_{ab}=\oomega_{[ab]}$, 
we obtain the following
expressions
\begin{align}
\omega_{c(ab)}&={1\over 2}\mathcal{Q}_{cab}, \nonumber \\[0.2cm]
\omega_{c[ab]}
&=\oomega_{cab}+K_{cab}+{1\over 2}\big(\mathcal{Q}_{abc}-\mathcal{Q}_{bca}\big).
\label{eq:connection_sym+antisym}
\end{align}
In particular, the metric-compatible connection
\begin{align}
\widetilde{\omega}_{bc}\equiv \omega_{[bc]},
\label{eq:omega_tilde_def}
\end{align}
has nonvanishing torsion given by (see, e.g.,~\cite{JimenezCano:2021rlu})
\begin{align}
\widetilde{T}^{a}&=T^{a}-{1\over 2}\mathcal{Q}^{a}_{\,\,\,b}\wedge \theta^{b}.
\label{eq:Ttilde_2form}
\end{align}
Under generalized 
metric rescalings~\eqref{eq:general_transform_frame} 
and~\eqref{eq:generalized_metric_rescalings_frame},~$\widetilde{\omega}^{a}_{\,\,\,b}$ 
transforms as
\begin{align}
\widetilde{\omega}^{a}_{\,\,\,b}\longrightarrow \widetilde{\omega}^{a}_{\,\,\,b}
+(1-\Lambda)\delta^{a}_{b}d\sigma,
\end{align}
being therefore Weyl invariant~($\Lambda=1$). Notice that
despite satisfying~$\widetilde{D}\eta_{ab}=0$, 
the connection~$\widetilde{\omega}_{ab}$ 
still has a dependence 
on the components of the original nonmetricity one-form~$\mathcal{Q}_{ab}$, as 
is seen from the second identity in eq.~\eqref{eq:connection_sym+antisym}. 
Let us mention, incidentally, that the nonmetricity-dependent
term in~\eqref{eq:Ttilde_2form} does not contribute to the torsional Chern-Simons 
form~\cite{Chandia:1997hu,Zanelli:2015pxa,Valle:2021nfv},~$\widetilde{\mathcal{H}}\equiv 
\theta_{a}\widetilde{T}^{a}=\theta_{a}T^{a}\equiv \mathcal{H}$, and therefore neither
to the corresponding Nieh-Yan invariant,~$\widetilde{N}\equiv d\widetilde{\mathcal{H}}=T_{a}\wedge T^{a}
-\theta^{a}\wedge\mathcal{R}_{ab}\wedge\theta^{b}\equiv N$.

\section{Generalized metric rescalings}
\label{sec:appendix}

In this appendix we review the implementation of generalized metric rescalings. 
We consider the family of transformations 
(cf.~\cite{Hehl:1994ue})
\begin{align}
\theta^{a}&\longrightarrow e^{\Lambda\sigma} \theta^{a}, \nonumber \\[0.2cm]
\eta_{ab}&\longrightarrow e^{2(1-\Lambda)\sigma}\eta_{ab}, 
\label{eq:general_transform_frame}\\[0.2cm]
\omega^{a}_{\,\,\,b}&\longrightarrow \omega^{a}_{\,\,\,b}
+(\Xi-\Lambda)\delta^{a}_{b}d\sigma,
\nonumber
\end{align}
with~$\Lambda$ and~$\Xi$ real parameters, while
the condition~$e_{b}\rfloor\theta^{a} =\delta^{a}_{b}$ implies
\begin{align}
e_{a}&\longrightarrow e^{-\Lambda\sigma}e_{a}.
\label{eq:transform_frame_e}
\end{align}
 The corresponding transformation of the
torsion, nonmetricity, and curvature forms read
\begin{align}
T^{a}&\longrightarrow e^{\Lambda\sigma}
\big(T^{a}+\Xi d\sigma\wedge \theta^{a}\big), \nonumber \\[0.2cm]
\mathcal{Q}^{a}_{\,\,\,b}
&\longrightarrow \mathcal{Q}^{a}_{\,\,\,b}+2\big(\Xi-1\big) \delta^{a}_{b}d\sigma,
\label{eq:generalized_metric_rescalings_frame} \\[0.2cm]
\mathcal{R}^{a}_{\,\,\,b}&\longrightarrow \mathcal{R}^{a}_{\,\,\,b},
\nonumber
\end{align}
while for the trace and traceless part of the nonmetricity tensor we find
\begin{align}
\mathcal{Q}&\longrightarrow \mathcal{Q}+2\big(\Xi-1\big)d\sigma, \nonumber \\[0.2cm]
\notrQ^{a}_{\,\,\,b}&\longrightarrow \notrQ^{a}_{\,\,\,b}.
\label{eq:notrQransf_Q_trans_general_app}
\end{align} 
Switching to spacetime indices, the transformations~\eqref{eq:general_transform_frame}
realize metric rescalings combined with
shifts in the linear connection, torsion, and nonmetricity
\begin{align}
g_{\mu\nu}&\longrightarrow e^{2\sigma}g_{\mu\nu}, \nonumber \\[0.2cm]
\Gamma_{\mu\nu}^{\,\,\,\,\,\,\lambda}&\longrightarrow 
\Gamma_{\mu\nu}^{\,\,\,\,\,\,\lambda}+\Xi\partial_{\mu}\sigma\delta^{\lambda}_{\nu}, 
\nonumber \\[0.2cm]
T_{\mu\nu}^{\,\,\,\,\,\,\lambda}&\longrightarrow 
T_{\mu\nu}^{\,\,\,\,\,\,\lambda}
+2\Xi\partial_{[\mu}\sigma \delta^{\lambda}_{\nu]},
\label{eq:general_transform_coordinates}\\[0.2cm]
\mathcal{Q}_{\mu\nu}^{\,\,\,\,\,\,\lambda}&\longrightarrow \mathcal{Q}_{\mu\nu}^{\,\,\,\,\,\,\lambda}
+2\big(\Xi-1\big)\partial_{\mu}\sigma\delta^{\lambda}_{\nu}.
\nonumber
\end{align}
The two nonmetricity vector fields introduced in 
eqs.~\eqref{eq:Q_Qbar_def} and~\eqref{eq:second_nm_gauge_field}
undergo the same gauge transformation
\begin{align}
\mathcal{Q}_{\mu}&\longrightarrow \mathcal{Q}_{\mu}+2\big(\Xi-1\big)\partial_{\mu}\sigma, \nonumber \\[0.2cm]
\widehat{\mathcal{Q}}_{\mu}&\longrightarrow \widehat{\mathcal{Q}}_{\mu}+2\big(\Xi-1\big)\partial_{\mu}\sigma.
\end{align}
In the language of ref.~\cite{Iosifidis:2018zwo},~$\Xi=0$ and~$\Xi=1$ correspond
to conformal transformations and frame rescalings, respectively. 
Following e.g.~\cite{Dereli:1982xb,Maluf:2011kf}, we refer
to the case~$\Lambda=1$, $\Xi=0$ 
as Weyl transformations.
``Pure gauge'' projective 
transformations~\cite{Iosifidis:2018zwo,Garcia-Parrado:2020lpt,Sauro:2022hoh,Rigouzzo:2023sbb} 
\begin{align}
g_{\mu\nu}&\longrightarrow g_{\mu\nu}, \nonumber \\[0.2cm]
\Gamma_{\mu\nu}^{\,\,\,\,\,\,\lambda}&\longrightarrow 
\Gamma_{\mu\nu}^{\,\,\,\,\,\,\lambda}+\partial_{\mu}\sigma'\delta^{\lambda}_{\nu},
\end{align}
are retrieved from previous expressions by setting~$\sigma=\Xi^{-1}\sigma'$ and
taking the limit~$\Xi\rightarrow\infty$.

Given a general~$r$-form tensor of type~$(p,q)$
of weight~$w$ under~\eqref{eq:general_transform_frame}
\begin{align}
\mathfrak{T}^{a_{1}\ldots a_{p}}_{\hspace*{0.8cm}b_{1}\ldots b_{q}}
\longrightarrow e^{w\sigma}\mathfrak{T}^{a_{1}\ldots a_{p}}_{\hspace*{0.8cm}b_{1}\ldots b_{q}},
\end{align}
a new covariant derivative can be defined for~$\Xi\neq 1$
\begin{align}
\mathscr{D}\mathfrak{T}^{a_{1}\ldots a_{p}}_{\hspace*{0.8cm}b_{1}\ldots b_{q}}
\equiv D\mathfrak{T}^{a_{1}\ldots a_{p}}_{\hspace*{0.8cm}b_{1}\ldots b_{q}}
+{(\Xi-\Lambda)(p-q)+w\over 2(1-\Xi)}
\mathcal{Q}\wedge \mathfrak{T}^{a_{1}\ldots a_{p}}_{\hspace*{0.8cm}b_{1}\ldots b_{q}},
\label{eq:Weyl_cov_der_def_new_param}
\end{align}
satisfying
\begin{align}
\mathscr{D}\mathfrak{T}^{a_{1}\ldots a_{p}}_{\hspace*{0.8cm}b_{1}\ldots b_{q}}
\longrightarrow e^{w\sigma}
\mathscr{D}\mathfrak{T}^{a_{1}\ldots a_{p}}_{\hspace*{0.8cm}b_{1}\ldots b_{q}}.
\end{align}
Particularizing to 
Weyl transformations ($\Lambda=1$, $\Xi=0$), the previous definition simplifies to
\begin{align}
\mathscr{D}\mathfrak{T}^{a_{1}\ldots a_{p}}_{\hspace*{0.8cm}b_{1}\ldots b_{q}}
\equiv D\mathfrak{T}^{a_{1}\ldots a_{p}}_{\hspace*{0.8cm}b_{1}\ldots b_{q}}
+{w-p+q\over 2}
\mathcal{Q}\wedge \mathfrak{T}^{a_{1}\ldots a_{p}}_{\hspace*{0.8cm}b_{1}\ldots b_{q}},
\label{eq:cov_der_def_new_param_Weyltrans}
\end{align}
in terms of which the torsion and shear nonmetricity are written as
\begin{align}
T^{a}&=\mathscr{D}e^{a}, \nonumber \\[0.2cm]
\notrQ_{ab}&=-\mathscr{D}\eta_{ab},
\end{align}
while the three Bianchi identities~\eqref{eq:Bianchi_id_overlineQ1} 
and~\eqref{eq:1st2dn_bianchi_id} take the form
\begin{align}
\mathscr{D}\notrQ_{ab}&=2\notrR_{(ab)}, \nonumber \\[0.2cm]
\mathscr{D}T^{a}&=\mathcal{R}^{a}_{\,\,\,b}\wedge e^{b}, 
\label{eq:bianchi_ids_calD}\\[0.2cm]
\mathscr{D}\mathcal{R}^{a}_{\,\,\,b}&=0.
\nonumber
\end{align}
Being expressed in terms of the covariant derivative
defined in~\eqref{eq:cov_der_def_new_param_Weyltrans}, 
the previous expressions are manifestly Weyl covariant.

\providecommand{\href}[2]{#2}\begingroup\raggedright\endgroup

\end{document}